\documentclass[twocolumn,floats,floatfix,amssymb,nofootinbib,prd,superscriptaddress,aps]{revtex4-2}
\usepackage{graphicx,amssymb,amsmath,amsthm,amsfonts,epsfig}
\usepackage[linktocpage,colorlinks=true,citecolor=OliveGreen,linkcolor=Maroon,urlcolor=Maroon]{hyperref}
\usepackage[usenames]{color}
\usepackage{epstopdf}
\usepackage{bm}
\usepackage{dcolumn}
\usepackage[utf8]{inputenc}
\usepackage{latexsym}
\usepackage{rotating}
\usepackage{hyperref}
\usepackage{tabularx}
\usepackage{braket}
\usepackage{color}
\usepackage{esvect}
\usepackage{enumerate}
\usepackage{footmisc}
\usepackage{array}
\usepackage{tensor}
\usepackage{mathtools}
\usepackage{url}
\usepackage[dvipsnames]{xcolor}
\usepackage{comment}
\usepackage{multirow}
\usepackage{nicematrix}
\usepackage{graphics}      
\usepackage[utf8]{inputenc} 
\usepackage{graphicx,amssymb,amsmath,amsthm,amsfonts,epstopdf,epsfig,epsf,times}
\usepackage[linktocpage]{hyperref}
\usepackage[usenames]{color}
\usepackage{textcomp}
\usepackage{bm}
\usepackage{latexsym}
\usepackage{rotating}
\usepackage{hyperref}
\usepackage{color}
\usepackage{enumerate}
\usepackage{tensor}
\usepackage{stmaryrd}
\usepackage[normalem]{ulem}
\usepackage{mathtools}
\usepackage{url}
\usepackage{multirow}
\usepackage{graphicx}
\usepackage{mathtools}
\usepackage{verbatim}
\usepackage{soul,xcolor}
\usepackage{xspace}
\usepackage{amsmath}
\usepackage{xfrac}
\usepackage{ragged2e}
\setstcolor{red}
\usepackage{esint}
\usepackage{physics}
\usepackage{subcaption}

\renewcommand{\vec}[1]{\boldsymbol{#1}}
\def\be{\begin{equation}}
\def\ee{\end{equation}}

\def\oc{\omega_c}

\newcommand{\beq}{\begin{eqnarray}}

\newcommand{\eeq}{\end{eqnarray}} 
\newcommand{\submu}{_{\mu}}

\newcommand{\supmu}{^{\mu}}

\newcommand{\mE}{\mathcal{E}}
\newcommand{\mL}{\mathcal{L}}
\newcommand{\lr}[1]{\left( #1\right)}

\newcommand{\qt}{\tilde{q}}

\def\be{\begin{equation}}
\def\ee{\end{equation}}
\def\bea{\begin{eqnarray}}
\def\eea{\end{eqnarray}}
\def\beq{\begin{eqnarray}}
\def\eeq{\end{eqnarray}}

\begin{document}

\title{Floating orbits and energy extraction from magnetized Kerr black holes}
\author{João S. Santos}
\affiliation{CENTRA, Departamento de F\'{\i}sica, Instituto Superior T\'ecnico -- IST, Universidade de Lisboa -- UL,
Avenida Rovisco Pais 1, 1049-001 Lisboa, Portugal}
\author{Vitor Cardoso}
\affiliation{CENTRA, Departamento de F\'{\i}sica, Instituto Superior T\'ecnico -- IST, Universidade de Lisboa -- UL,
Avenida Rovisco Pais 1, 1049-001 Lisboa, Portugal}
\affiliation{Niels Bohr International Academy, Niels Bohr Institute, Blegdamsvej 17, 2100 Copenhagen, Denmark}
\author{José Natário}
\affiliation{CAMSDG, Departamento de Matem\'{a}tica, Instituto Superior T\'ecnico -- IST, Universidade de Lisboa -- UL,
Avenida Rovisco Pais 1, 1049-001 Lisboa, Portugal}

\date{\today}

\begin{abstract}
We study the electromagnetic radiation reaction on a charged particle around a weakly magnetized Kerr black hole, 
by numerically solving the Teukolsky equation to find the energy fluxes in electromagnetic radiation at the horizon and at spatial infinity. We also employ analytical methods in the low-frequency limit, finding excellent agreement with the numerical results. For a wide range of parameters, the energy fluxes on the horizon are negative for all orbital radii: the modes are amplified through superradiance. More interestingly, the flux on the horizon is larger (in absolute value) than the flux at infinity for orbits with orbital radius $r_0 \gtrsim 9M$ --  \textit{floating orbits} are a generic outcome of having magnetic fields around black holes. Particles on these orbits extract energy from the black hole through their radiation field; the net effect is an increase in the particles' kinetic energy, so that they outspiral to infinity. Although in realistic accretion disks viscous forces are expected to dominate, floating orbits remain an interesting feature of the existence of ergoregions in black hole spacetimes.
\end{abstract}

\maketitle
\section{Introduction} \label{sec:intro}
%
The Kerr solution to the vacuum Einstein field equations~\cite{Kerr:1963ud} describes the unique stationary and asymptotically flat black hole (BH) solution in General Relativity~\cite{Hawking:1971vc,  PhysRevLett.26.331, PhysRevLett.34.905, Chrusciel:2012jk}. It represents an isolated rotating BH, and is characterized by two parameters only -- the mass and the angular momentum. This is the simplest theoretical description of any macroscopic object known to date!

In the last decade, there has been mounting experimental evidence pointing to BHs being commonplace in the Universe. Namely, we refer to the observations of gravitational waves from binary BH coalescences in the LIGO-Virgo-Kagra observatories~\cite{LIGOScientific:2018mvr,LIGOScientific:2020ibl, KAGRA:2021vkt}, and of supermassive BHs at the center of galaxies~\cite{GRAVITY:2020gka,EventHorizonTelescope:2019dse,EventHorizonTelescope:2022wkp}. So far, all observations are compatible with the dark compact objects being Kerr BHs described by General Relativity. The ``Kerr paradigm''~\cite{Cardoso:2019rvt} will be tested with increasing precision in the near future with space-based~\cite{LISA:2017pwj} and other next generation detectors~\cite{Maggiore:2019uih,Reitze:2019iox,TianQin:2015yph}. Thus, the study of the rich physics occurring in the vicinity of BHs is an active and timely topic.

Astrophysical BHs are not in isolation: instead, they are immersed in complex and dynamic systems~\cite{GRAVITY:2020gka,EventHorizonTelescope:2019dse,EventHorizonTelescope:2021bee,EventHorizonTelescope:2019pgp,EventHorizonTelescope:2021srq,EventHorizonTelescope:2022wkp}. For example, BHs may be surrounded by an accretion disk of ionized matter~\cite{EventHorizonTelescope:2021bee,EventHorizonTelescope:2019pgp,EventHorizonTelescope:2021srq,Tanaka:1995en,Page:1974he,Thorne:1974ve,Abramowicz:2011xu}, which can support appreciably strong magnetic fields. Indeed, recent observation show that BHs are commonly eveloped by magnetic fields of order $\sim10^4$ and $\sim10^8$ Gauss, for supermassive and stellar mass BHs, respectively~\cite{2011AstBu..66..320P,Eatough:2013nva,Baczko:2016opl,Daly:2019srb, EventHorizonTelescope:2021srq}. 
Here, we consider a weak magnetic field, asymptotically uniform and aligned with the BH axis of rotation. We take the magnetic field weak enough that the geometry is described by the Kerr metric~\cite{PhysRevD.10.1680,Ipser:1971zz}. This zero backreaction approximation is valid only if we restrict to a region of spacetime satisfying~\cite{Galtsov:1978ag}
\begin{equation}
	\frac{r }{G M / c^2} \ll   4.7\times 10^{19} \lr{\frac{1\, \text{Gauss}}{B_0}}\lr{\frac{M_\odot}{M}} \, ,\label{eq:limit_validity}
\end{equation}
with $G$ Newton's gravitational constant, $c$ the speed of light and $M_\odot$ the the mass of the Sun. For astrophysical magnetic fields around BHs~\cite{2011AstBu..66..320P,Eatough:2013nva,Baczko:2016opl,Daly:2019srb, EventHorizonTelescope:2021srq}, this approximation has a very wide region of applicability. 


Kerr BHs are prone to energy extraction mechanisms~\cite{Penrose:1969pc,1972JETP...35.1085Z,Starobinsky:1973aij,Starobinskil:1974nkd,Brito:2015oca,PhysRevD.16.1615,PhysRevLett.114.251103,Ruffini:1975ne,1977MNRAS.179..433B}, which convert the BH rotational energy into kinetic energy of particles or occupation number of fields. In this paper we report an interesting effect of superradiance~\cite{Brito:2015oca}, whereby a charged particle orbiting a weakly magnetized Kerr BH can extract rotational energy from the latter and {\it outspiral}. This finding is another example of how magnetic fields can provide means of energy extraction from BHs~\cite{Ruffini:1975ne,1977MNRAS.179..433B,1985ApJ...290...12W,1986ApJ...307...38P,1986ApJ...301.1018W,Chakraborty:2024aug,Gupta:2021vww,Kolos:2020gdc,Brito:2014nja,Rueda:2023mtp,Comisso:2020ykg}.
In a previous paper~\cite{Santos:2024tlt}, we studied the the effect of radiation reaction in the weakly magnetized Schwarzschild system. We found a ``horizon dominance effect" for a specific orbital configuration -- the plus configuration. In this work, we will look at a similar orbital configuration -- the prograde plus configuration -- and show that the radiation emitted by the particle is now amplified through superradiance. This amplification is strong enough 
for the particle to experience a net energy gain and outspiral to infinity, following a so-called \emph{floating orbit}. This type of orbit cannot be found in the unmagnetized Kerr system~\cite{Torres:2020fye,German:2023bye}, unless massive fundamental fields are invoked~\cite{Cardoso:2011xi}. 

We focus on charged particles in equatorial-plane circular orbits around weakly magnetized Kerr BHs. We use the Teukolsky formalism~\cite{Teukolsky:1972my,Teukolsky:1973ha} to obtain the electromagnetic (EM) energy and angular momentum fluxes on the horizon and at infinity~\cite{Teukolsky:1974yv,Hawking:1972hy, Poisson:1993vp, Poisson:1994yf}. By energy and angular momentum conservation, knowledge of the fluxes allows one to evolve the orbital parameters and study particle trajectories, as is usually done in the absence of magnetic fields~\cite{Poisson:1993vp,Poisson:1994yf,German:2023bye,Cardoso:2011xi,Yunes:2011aa,Lynch:2021ogr}. Although this orbital evolution is not studied in detail here, we show in Appendix~\ref{app:adiabatic_approximation} that particles in circular orbit do remain in circular orbit when driven by radiation reaction. Note that this calculation will only capture dissipative effects, even though the full self-force also includes a (much smaller) conservative counterpart~\cite{Smith:1980tv,Poisson:2011nh,Barack:2018yvs}.

This paper is organized as follows: in Sec.~\ref{sec:background} we intoduce the weakly magnetized Kerr BH and look at the various types of circular orbits in this system. In Sec.~\ref{sec:perturbations} we use the Teukolsky equation to obtain a multipolar expansion for energy and angular momentum fluxes, and in Sec.~\ref{sec:results} we present and discuss our analytical and numerical results. Finally, in Sec.~\ref{sec:conclusions} we make some concluding remarks and comment on prospects for future work. 

We use geometric units $G=c=1$, and a Gaussian electromagnetic unit system with $ 4\pi \varepsilon_0 = \mu_0 /4 \pi =1 $. The metric signature is $(-,\, +,\, +,\, +)$, and Greek indices run from $0$ to $3$.
%
%
%
\section{Setup} \label{sec:background}
%
%
\subsection{Weakly magnetized Kerr BH} \label{sec:mag_bh}
%
\begin{figure*}[ht!]
\centering
\begin{subfigure}{0.22\textwidth}
\centering
\includegraphics[width=\textwidth]{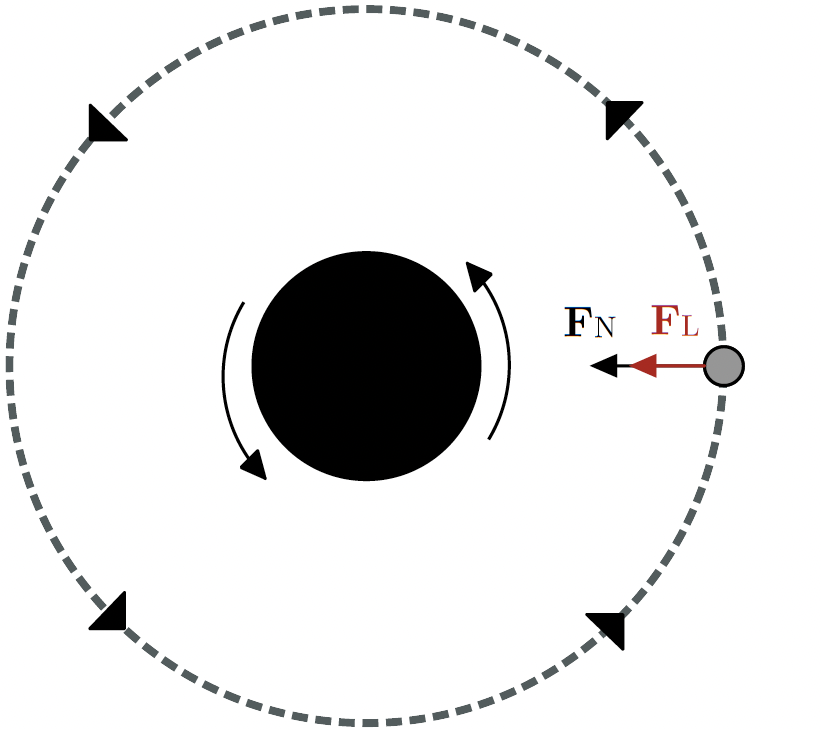}
\caption{Prograde minus : $a>0 \, , \ \oc < 0$.}
\end{subfigure}
\hspace{0.02 \textwidth}
\begin{subfigure}{0.22\textwidth}
\centering
\includegraphics[width=\textwidth]{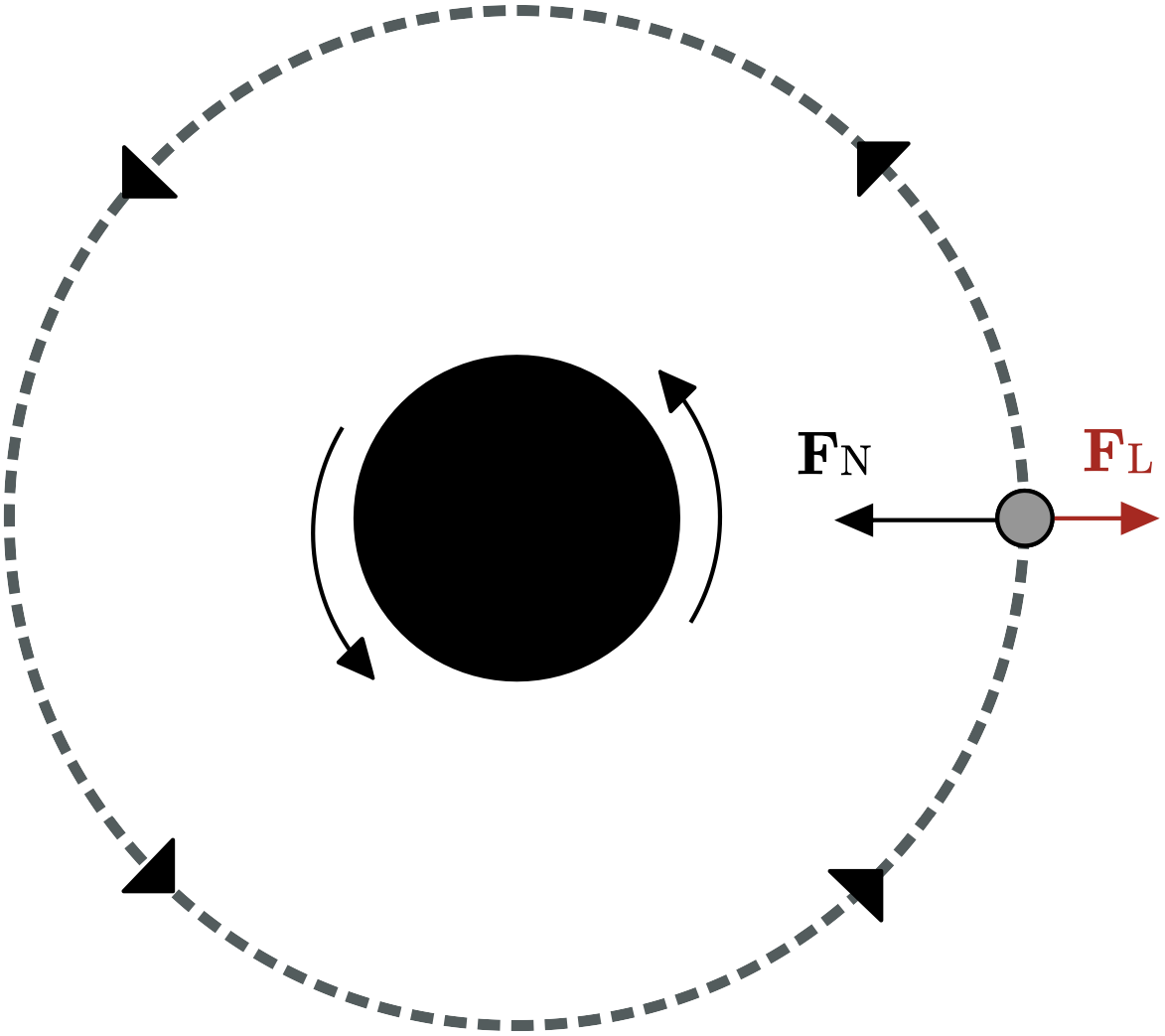}
\caption{Prograde plus : $a>0 \, , \ \oc > 0$.}
\end{subfigure}
\hspace{0.02 \textwidth}
\begin{subfigure}{0.22\textwidth}
\centering
\includegraphics[width=\textwidth]{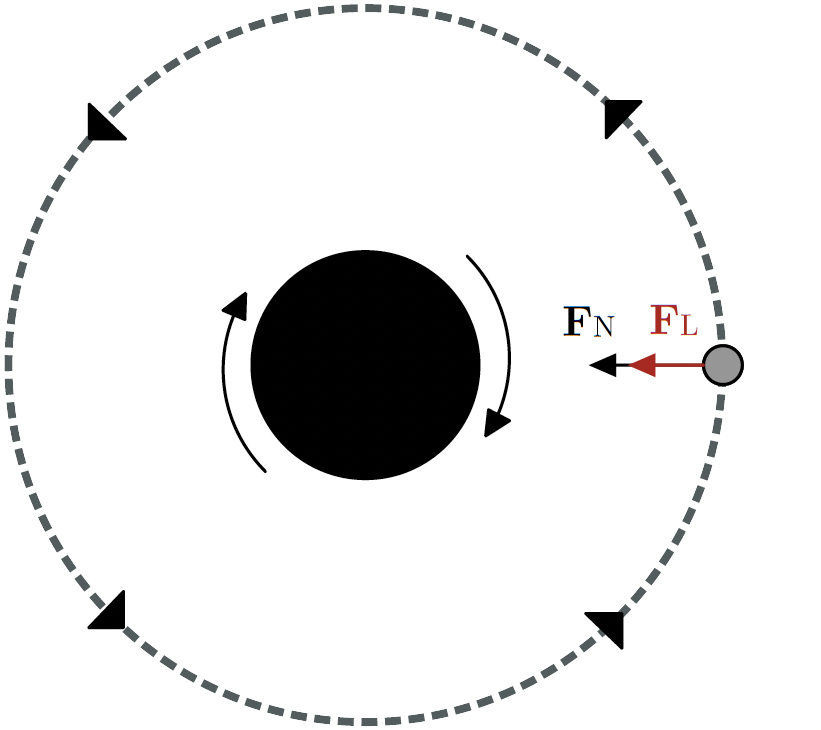}
\caption{Retrograde minus : $a<0 \, , \ \oc < 0$.}
\end{subfigure}
\hspace{0.02 \textwidth}
\begin{subfigure}{0.22\textwidth}
\centering
\includegraphics[width=\textwidth]{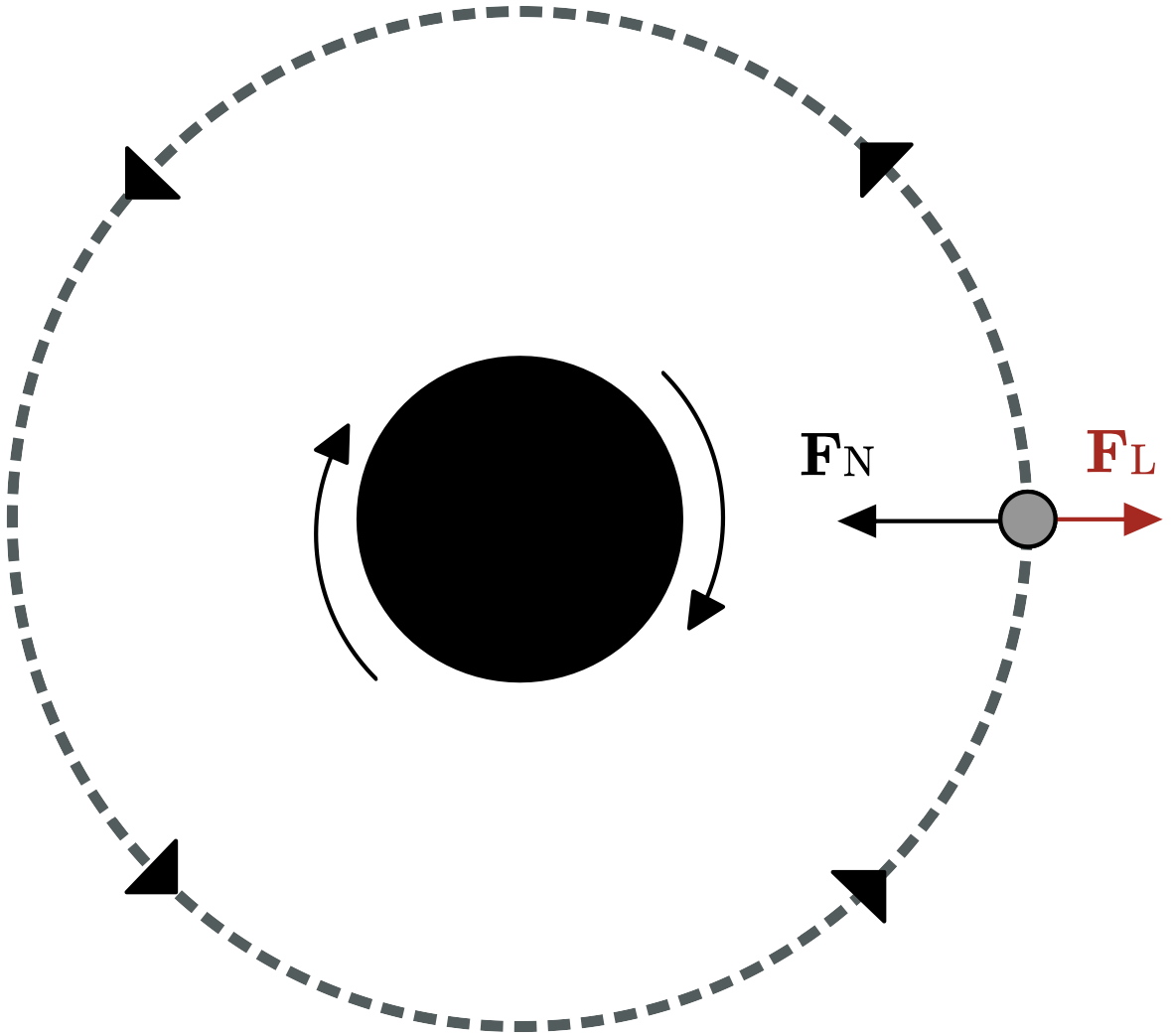}
\caption{Retrograde plus : $a<0 \, , \ \oc > 0$.}
\end{subfigure}
\caption{\justifying Schematic representation of the four configurations of equatorial circular orbits in the weakly magnetized Kerr BH system. The BH is immersed in an asymptotically uniform magnetic field; both the BH angular momentum and the asymptotic magnetic field are orthogonal to the plane of this page. The charged particle (grey dot) is assumed to orbit the BH (black disk at the center) anticlockwise; the BH rotation, parametrized by $a$, is indicated by black arrows in the center; the cyclotron frequency is $\oc = q B_0 / m$, where $B_0$ is the magnetic field strength at infinity, $q$ is the charge of the particle, and $m$ is its mass. The configurations are invariant under reflections about the plane of the paper. We also include a Newtonian interpretation of the difference between the plus and minus configurations: $\vec{F}_\text{N}$ and $\vec{F}_\text{L}$ are, respectively, the gravitational and Lorentz forces; the latter is centripetal (centrifugal) in minus (plus) configuration orbits.}\label{fig:configurations}
\end{figure*}
The Kerr family of solutions is described by two parameters: the mass $M$ and spin $a$. In Boyer Lindquist coordinates coordinates $\{t,r,\theta,\phi \}$, the corresponding line element is
\beq
ds^2 &=& - \left(1-\frac{2Mr}{\Sigma}\right) dt^2 - \frac{4 M a r \sin^2 \theta}{\Sigma} dt d\phi+\frac{\Sigma}{\Delta} dr^2  \nonumber \\ 
& +& \Sigma d\theta^2 + (r^2 +a^2 
+ \frac{2 M r a^2 }{\Sigma} \sin^2 \theta) \sin^2 \theta \; d\phi^2 \, , \label{eq:kerr_metric}
\eeq
where
\be
\Sigma =r^2 + a^2 \cos^2 \theta \, , \quad \Delta = r^2 -2 M r + a^2
\ee
and $|a| \leq M$. The inner and outer horizons are located at the roots of $\Delta$,
\begin{equation}
r_\pm = M \pm \sqrt{M^2 -a^2}\, .
\end{equation}
The outer horizon at $r_+$ is the event horizon of the BH. This metric is stationary and axisymmetric, admitting the two Killing vector fields
\begin{equation}
	X^\mu = \delta ^\mu _t \qquad \text{and} \qquad Y^\mu = \delta ^\mu _ \phi \, , 
	\label{eq:Killing_vectors}
\end{equation}
where $\delta^\mu _\nu$ is the identity operator. These vectors are associated with conservation laws for energy and angular momentum along $\hat{z}$, which we define in Eq.~\eqref{eq:conserved_qtys}.

A free particle of mass $m$ will follow timelike geodesics. One particular family of geodesics are the circular orbits in the equatorial plane ($\theta = \pi / 2$), characterized by the orbital radius $r_0$ and the orbital frequency measured by asymptotic stationary observers, $\Omega_0 = d \phi/ dt$. The condition that the circular orbit is a geodesic of the Kerr geometry is
\begin{equation}
    \Omega_0 = \omega_{p,r} =  \frac{\pm\sqrt{M}}{r_0 ^{3/2} \pm a \sqrt{M}} \, , \label{eq:omega_geo}
\end{equation}
where plus (minus) sign is assigned to $\omega_p$ ($\omega_r$), which corresponds to prograde or co-rotating (retrograde or counter-rotating) motion~\cite{Bardeen:1972fi}.

In the fixed Kerr geometry~\eqref{eq:kerr_metric}, we consider a test (no backreaction) EM field. For a stationary and axisymmetric field that asymptotes to a uniform magnetic field along the $\hat{z}$ direction, the vector potential~\cite{PhysRevD.10.1680,Ipser:1971zz} is
\begin{align}
A\submu =& \frac{B_0}{2} \, Y\submu \, , \label{eq:mag_potential}
\end{align}
where $\vec{B} = B_0 \hat{z}$ is the asymptotic magnetic field. Recall that  the test field approximation requires such a solution to be restricted to  values of $r$ satisfying Eq.~\eqref{eq:limit_validity}.
\subsection{Circular orbit motion}
Consider now a test particle of mass $m$, and electric charge $q$ moving in this system. We can write the Hamiltonian as
\be
H \left(\pi_\mu ,  x\supmu\right) =  \frac{1}{2} \left[g^{rr} \pi_r ^2 + g^{\theta\theta} \pi_\theta ^2 + m^2 \mathcal{F} \right]\, , \label{eq:sch_charged_hamiltonian}
\ee
where $\pi_\mu= m u_\mu + q A_\mu$ are the conjugate momenta, with $u^\mu=dx^\mu / d\tau$ the particle's 4-velocity and $\tau$ the proper time, and $\mathcal{F}=\mathcal{F}(r,\theta;\mE,\mL,\oc)$ is defined from the energy per unit mass $\mE$ and the angular momentum along $\hat{z}$ per unit mass $\mL$,
\begin{equation}
	\begin{split}
		&\mE = \frac{E}{m} = - \frac{\pi_t}{m} = -\left(u_t + \qt A_t \right),  \\ 
		&\mL = \frac{L}{m} = \frac{\pi_\phi}{m} = \left(u_\phi + \qt A_\phi  \right)
	\end{split}
 \label{eq:conserved_qtys}
\end{equation}
(both conserved along the particle worldline) as
\beq
\mathcal{F} &= &1 + g^{tt}\left(\mE + \frac{\oc}{2}  g_{t\phi} \right)^2 + g^{\phi \phi}\left(\mL - \frac{\oc}{2}g_{\phi\phi} \right)^2 \label{eq:FF} \\
&-&2 g^{t\phi}\left(\mE + \frac{\oc}{2} g_{t\phi} \right)\left(\mL - \frac{\oc}{2} g_{\phi\phi} \right) \,.\nonumber 
\eeq
In what follows, instead of the magnetic field strength, we characterize the magnetic field using the (possibly negative) cyclotron frequency of the particle, 
\begin{equation}
	\omega_c \equiv \frac{q B_0}{m} \, .
	\label{eq:oc}
\end{equation}
In fact, in our analysis, we will only be looking at ratios of energy fluxes, so the results really depend only on $\oc$\footnote{However, if one wishes to actually construct the trajectory of the particle then the three parameters $q$, $B_0$ and $m$ have to be specified.}.

We can use the Hamilton equations of motion to study the general motion of particles in the magnetized Kerr spacetime. Numerical studies of the general motion were conducted in~\cite{Aliev:2002nw,Shiose:2014bqa,Zhang:2017nhl}. Here, we focus on circular orbits of radius $r_0$ in the equatorial plane, which can be found by solving 
\beq
\mathcal{F} =  \partial_r \mathcal{F} = 0 \, , \label{eq:circular_cond}
\eeq
where we note that the equation $\partial_\theta \mathcal{F} = 0$ is trivially satisfied if the the other two hold. A given circular orbit is stable if 
\be
\partial_r ^2\mathcal{F}>0 \, ,\quad\partial_\theta ^2\mathcal{F}>0 \, .
\ee
This system exhibits a reflection symmetry, given by the transformation $(a, \mL, \oc)\to ( -a,-\mL, -\oc)$, and so we can always assume $\mL>0$. Consequently, circular orbits can be divided into four different configurations:

\noindent \emph{Prograde minus configuration}: $a>0$ and $\oc< 0$. 

\noindent \emph{Prograde plus configuration}:  $a>0$ and $\oc> 0$. 

\noindent \emph{Retrograde minus configuration}: $a<0$ and $\oc> 0$. 

\noindent \emph{Retrograde plus configuration}: $a<0$ and $\oc< 0$. 

These are represented in Fig.~\ref{fig:configurations}. Each of these configurations is characterized by different dependencies of the orbital frequency $\Omega_0 = d\phi / dt$ on the radius $r_0$. 

We obtained a full characterization of the two configurations present in the case of a magnetized Schwarzschild BH~\cite{Santos:2024tlt}, where there is no distinction between prograde and retrograde orbits. The Kerr case is qualitatively similar \cite{Tursunov:2016dss}, the only difference being the orbital frequency of particles following geodesics, which is now given by Eq.~\eqref{eq:omega_geo}.

In the study of the magnetized Schwarzschild BH~\cite{Santos:2024tlt}, we found a ``horizon dominance effect" in plus configuration orbits: for sufficiently large $M \oc$, the energy flux on the horizon dominates the energy flux at infinity, even for arbitrarily wide circular orbits. This effect occurs for low orbital frequency $M \Omega_0 \ll 1$, so that the radiation emitted is also of low frequency. Thus, when the BH is rotating, these modes are candidates for superradiant amplification~\cite{Brito:2015oca}. In particular, for the case of particles in circular orbits, this effect only occurs if the particle is corrotating (cf.~Eq.~\eqref{eq:superradiance}). Although the high frequency prograde minus configuration orbits also stimulate superradiance, the amplification is not strong enough to lead to floating orbits, and so we focus exclusively on prograde plus configuration orbits.

In the prograde plus configuration, we can distinguish two asymptotic regimes, separated by a critical radius, 
\begin{equation}
    r_c = \left( \frac{M}{\oc^2}\right)^{1/3} \, , \label{eq:rc}
\end{equation}
just as in the case of Schwarzschild~\cite{Santos:2024tlt}. The asymptotic regimes are then
\begin{equation}
    \Omega_0 \sim \omega_p \quad (r_0 \ll r_c) \ \  , \qquad   \Omega_0 \sim \frac{\omega_p^2}{\omega_c}  \quad (r_0 \ll r_c) \label{eq:omega_PC} \, . 
\end{equation}
We will only study one particular set of parameters in this work, namely
\beq
a = 0.7 M \, , \quad M \oc = 0.1 \, .
\label{eq:parameters}
\eeq
These are motivated by gravitational wave observations of nearly equal mass mergers~\cite{LIGOScientific:2020tif,LIGOScientific:2021sio}. The value of $\oc$ can  be as high as $M \omega_c \sim 10^{11}$ for an electron orbiting a supermassive BH, according to recent measurements of the magnetic fields surrounding the latter~\cite{2011AstBu..66..320P,Eatough:2013nva,Baczko:2016opl,Daly:2019srb}. We chose a more conservative value due to limited numerical precision,  and also because for large values of $M \oc$ circular orbits become unstable under perturbations orthogonal to the equatorial plane~\cite{Tursunov:2016dss}. Still, it is important to note that for different values of $a$ and $\oc$ we found qualitatively similar results. 

\begin{figure}
    \centering
    \includegraphics[width= .5 \textwidth]{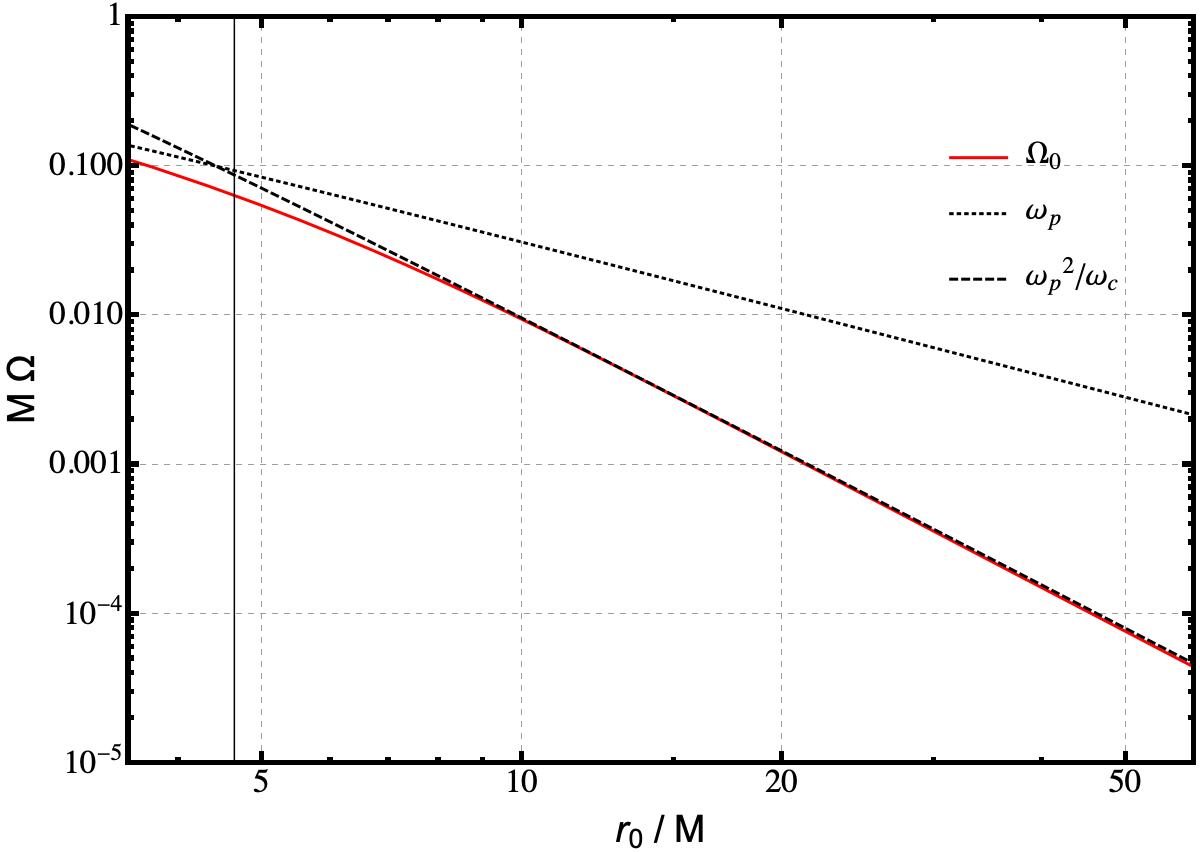}
    \caption{\justifying  Profile of orbital velocity $\Omega_0=d \phi/dt$, for prograde plus circular orbits of radius $r_0$ around a weakly magnetized BH with $a=0.7 M$ and $M \oc = 0.1 $. The red curve corresponds to the numerically calculated orbital velocity profile. The black dotted curve is the orbital frequency for prograde geodesics, see Eq.~\eqref{eq:omega_geo}; the black dashed curve is obtained using Eq.~\eqref{eq:omega_PC}. Lastly, the black vertical line indicates the critical radius $r_c \approx 4.64 M$ (see Eq.~\eqref{eq:rc}). This result confirms the asymptotic behaviour predicted in Eq.~\eqref{eq:omega_PC}.} 
    \label{fig:vel_prof}
\end{figure}

We will therefore focus on prograde plus configuration orbits for the system parameters shown in Eq.~\eqref{eq:parameters}. The orbital velocity profile is shown in Fig.~\ref{fig:vel_prof} for radii outside  the innermost circular orbit (ISCO), which we find to be 
\begin{equation}
    r_\text{ISCO} \approx 3.13 M \, . \label{eq:ISCO}
\end{equation}
%
%
%
%
\section{Radiation reaction in Teukolsky formalism} \label{sec:perturbations}
%
We now want to introduce radiation emission. We do so by using BH perturbation theory~\cite{Regge:1957td,Zerilli:1970se}, namely the Teukolsky equation~\cite{doi:10.1063/1.1724257,Teukolsky:1972my,Teukolsky:1973ha}. This formalism allows us to calculate the radiation field, and from it the energy fluxes on the horizon and at infinity. We will give only an outline of the calculation here, since it follows the same procedure used for non-rotating BHs in~\cite{Santos:2024tlt}.
%
\subsection{The Teukolsky equation}
The Teukolsky equation describes the radiative degrees of freedom of a spin-$s$ massless field~\cite{Teukolsky:1973ha}, encapsulated in a master variable $\psi$ sourced by matter fields $T$. The master equation is separable into a radial and an angular part by expanding
\begin{equation}
	\psi = \int _{- \infty} ^\infty d \omega \sum _{\ell,m } ^\infty R_{\omega \ell m } (r) _s
 S_{\omega \ell} (\theta) e^{i m \phi} e^{-i \omega t} \, ,\label{eq:psi_fourier_harmonic}
\end{equation}
and similarly for the source $T$,
%
where the functions $_sS_{\omega \ell} (\theta)$ are the spin weighted spheroidal harmonics~\cite{Press:1973zz,1986JMP....27.1238L}. Separation of variables yields two ordinary differential equations, one in the radial coordinate $r$ and another in the $\theta$ coordinate. The angular equation is a regular Sturm-Liouville problem with eigenvalues $_s E_{\omega \ell}$, and defines spin-weighted spheroidal harmonics. The radial equation is a Schrödinger-like equation;  it takes the form, omitting the subscripts in $R$ and $T$ for simplicity,
\beq
&&\Delta^{-s} \partial_r \lr{\Delta^{s+1}\partial_r R} - V(r) R = - 4 \pi \Sigma T\,,\label{eq:teukolsky_radial}\\
&&V(r) = \frac{K^2 -2is (r-M) K}{\Delta} + 4 i s \omega r - \lambda \, , \nonumber
\eeq
where $K= (r^2 + a^2)\omega - a m  $ and $\lambda= {_s E_{\omega \ell}} - 2 a m \omega + a^2 \omega^2 - s(s+1)$. There is no closed form analytical expression for the eigenvalues or eigenfunctions of the angular equation, so they have to be computed numerically~\cite{Press:1973zz,1986JMP....27.1238L,Berti:2005gp}. We use the package \texttt{SpinWeightedSpheroidal} of the Black Hole Perturbation Toolkit~\cite{BHPToolkit}.

Thus, the problem amounts to solving the radial equation~\eqref{eq:teukolsky_radial} with appropriate boundary conditions, corresponding to purely ingoing waves ot the horizon and purely outgoing waves at infinity. This is done by first constructing two linearly independent solutions -- $R^H$ and $R^\infty$ -- to the homogeneous problem ($T=0$), such that they satisfy the intended boundary conditions at the horizon and at infinity, respectively. A straightforward calculation~\cite{Teukolsky:1973ha} reveals that these solutions must have the following asymptotics:
\beq
R^H &\sim & A_{\text{in}}\  r^{-1}\ e^{- i \omega r_\star} + A_{\text{out}} \ r^{-2s-1} \ e^{ i \omega r_\star} \, , \nonumber \\
R^\infty &\sim&  r^{-2s-1} \ e^{i \omega r_\star}\qquad (r\to\infty)\,, \label{eq:asymptotics_inf} \\ \vspace{0.5 cm}
R^\infty &\sim & B_{\text{in}} \ \Delta^{-s} e^{- i k r_\star} + B_{\text{out}}\ e^{ i k r_\star} \, ,  \nonumber \\
R^H & \sim & \Delta^{-s} \ e^{- i k r_\star}\qquad (r\to r_+)\,, \label{eq:asymptotics_hor}
\eeq
where $r_\star$ is the usual tortoise coordinate in the Kerr metric and $k = \omega - m \Omega_H$, for $\Omega_H = a/2 M r_+$ the angular velocity of the Kerr BH's horizon~\cite{Teukolsky:1973ha}. Note that if 
\begin{equation}
    k<0 \iff m \Omega_H/\omega > 1 \,, \label{eq:superradiance}
\end{equation} 
wave amplification (superradiance) at the BH horizon is possible~\cite{Brito:2015oca}. The low orbital frequency of the prograde plus configuration makes it especially prone to superradiant amplification. 

Going back to the problem of finding solutions to the Teukolsky equation, once we find $R^H$ and $R^\infty$, we can then construct a rescaled  Wronskian
\begin{equation}
\mathcal{W}=\Delta^{s+1} \lr{R^\infty \frac{d R^H 
}{d r}-R^H \frac{d R^\infty}{d r}}\, . \label{eq:Wronskian}
\end{equation}
The solution to the non-homogeneous problem ($T\neq0$) is then simply
\beq
R(r) &=& \frac{R ^\infty (r)}{\mathcal{W}}  \int_{r_+} ^r R ^H (r^\prime) T(r^\prime) \ \Delta^s d r^\prime \nonumber\\
&+& \frac{R ^H (r)}{\mathcal{W}} \int_{r} ^\infty R ^\infty (r^\prime) T (r^\prime) \ \Delta^s d r^\prime \, . \label{eq:gen_sol_r}
\eeq
%

\subsection{EM radiation by charged particle in a circular orbit} \label{sec:EM_energy_circular_orbit}
%
For EM fields, energy fluxes can be computed from a single scalar quantity, $\phi_2$, which is a particular projection of the Faraday tensor onto the Kinnersley tetrad with $s=-1$. This quantity can be found by solving the master equation (Eq.~4.7 of Ref.~\cite{Teukolsky:1973ha}) with
\begin{align}
\psi = & (r-i a \cos \theta)^2 \phi_2 \, , \quad 
T =  (r-i a \cos \theta)^2 J_2\,, \label{eq:sub_teuk}
\end{align}
where $J_2$ is defined in Eq.~3.8 of Ref.~\cite{Teukolsky:1973ha}. The functional form of the EM 4-current $J_\mu$ is the same as in the case where the background is the Schwarzschild solution, and reads~\cite{Santos:2024tlt}
\begin{equation}
	J_\mu (x) = q \frac{u_\mu}{u^t} \frac{\delta (r-r_0)}{r^2} \frac{\delta(\cos \theta)}{\sin \theta} \delta(\phi - \Omega_0 t)\,.
\end{equation}
Plugging these expressions into Eq.~\eqref{eq:gen_sol_r} yields  
\begin{widetext}
\beq
\phi_2 &=& \frac{1}{(r-i a \cos \theta)^2} \sum_{\ell, m} \left[ \left(R^\infty (r) \, Z^\infty _{\ell m} \, \Theta(r-r_0) +  R^H(r) \, Z^H _{\ell m}\,\Theta(r_0-r)\right) \ _{-1} S_{\omega  \ell } (\theta) \, e^{i (m \phi - \omega t)} \right]\, ,\label{eq:phi2} \\
Z^{\infty,H}_{\ell m}&\equiv&   \frac{4 \pi q}{\mathcal{W}} \Bigg[ \frac{{_{-1}S_{\omega \ell }} }{2 \sqrt{2} \, r_0^3}  \left(-i a (g_{t t}+g_{t \phi} \Omega_0)-i (g_{t \phi}+g_{\phi \phi}\Omega_0)\right) \nonumber \\
&&  \times \left( \left( \frac{i \, \omega \left(r_0^2 + a^2 \right)-i \, a \, m}{ \Delta} +\frac{3 }{r_0} \right) r_0 ^2 \, R^{H,\infty} -\frac{d}{dr} \lr{r^2 R^{H, \infty}} \right) \nonumber \\
&& - \frac{R^{H,\infty}}{2 \sqrt{2} \, r_0^4 \, \Delta} \left( \lr{r_0 ^2 +a^2} (g_{t t}+g_{t \phi} \Omega_0) +a (g_{t \phi}+g_{\phi \phi} \Omega_0)\right)\nonumber \\
&& \times \left( \left( a \, \omega - m - \frac{ 2 \, i \, a \, r_0 }{ r_0^2 } \right) r_0 ^3 {_{-1}S_{\omega \ell }} - \frac{d}{d\theta} \big( {_{-1}S_{\omega \ell }} \, \Sigma \, (r_0- i \,a \cos \theta ) \sin \theta 
 \big)\right) \Bigg]  \, , \label{eq:Zlm}	
\eeq
\end{widetext}
where $\omega =m\Omega_0$, $\Theta(\cdot)$ is the Heaviside step function, $g_{\mu \nu}$ are the components of the Kerr metric~\eqref{eq:kerr_metric}, and all quantities in $Z^{\infty,H} _{\ell m}$ are evaluated at $r=r_0$, $\theta=\pi/2$; recall that the functions $R^{\infty, H}$ carry subscripts $\omega \ell m$. 

Finally, the energy fluxes at infinity and at the horizon can be easily computed~\cite{Teukolsky:1974yv}:
\begin{equation}
	\dot{E}^\infty  \equiv \frac{d E}{dt}\Bigg|_{\infty} = \frac{1}{2 \pi} \sum_{\ell ,m} \left| Z^\infty _{\ell m} \right|^2 \, ; 
 \label{eq:ene_flux_inf}
\end{equation}
\begin{equation}
\dot{E}^H = \sum_{\ell ,m} \frac{64 \, \omega \, k \, M^3 r_+^3 (k^2 + 4 \epsilon)}{\pi B^2}  \left| Z^H  _{\ell m} \right| ^2 \, ,\label{eq:ene_flux_hor} 
\end{equation}
where
$$\epsilon = \frac{\sqrt{M^2 - a^2}}{4 M r_+}$$
and
$$B^2 = \left(a^2 \omega ^2- 2 \, a \, m \, \omega + {_{-1}E_{\omega \ell}}\right)^2-4 \, a^2 \omega ^2+4 \, a \, m  \, \omega \, .$$
From these expressions, it is trivial to obtain angular momentum fluxes, as they are generally related to energy fluxes by ~\cite{Bekenstein:1973mi, Teukolsky:1974yv,Cardoso:2020iji,German:2023bye} 
\begin{equation}
\dot{L}^{\infty,H} = \frac{m}{\omega} \dot{E}^{\infty,H}\, .\label{eq:flat_momentum_change}
\end{equation}
Thus, once the homogeneous radial Eq.~\eqref{eq:teukolsky_radial} is solved, we can determine the full solution. In particular, from $R^{\infty,H}$ we can determine the energy and angular momentum fluxes in a straightforward manner. We will solve the homogeneous equation by employing both analytical and numerical methods. 
\subsection{Comment on adiabatic evolution} \label{sec:comment_adiabatic}
%
The motion of a particle in the equatorial plane of this system is characterized by two quantities -- the energy and angular momentum defined in Eq.~\eqref{eq:conserved_qtys}. For circular motion, a single parameter suffices, which we choose to be the energy. 

Our results can be used to study the motion of a charged particle around a weakly magnetized Kerr BH, by taking a sequence of circular orbits, parameterized by energy. Since energy is globally conserved, the particle's energy evolves as 
\begin{equation}
\dot{E}=-\dot{E}^\infty-\dot{E}^H \, . 
\end{equation}
This is the adiabatic approximation~\cite{Isoyama:2021jjd,German:2023bye,Skoupy:2023lih}, valid if the radiation timescale is much larger than an orbital period, i.e. for
\begin{equation}
\left|\dot{E} / E \right| \ll \Omega_0 \, . \ 
\end{equation}  

On the other hand, even within the adiabatic regime, it is possible that radiation renders circular motion unstable. Conversely, if one slightly perturbs an orbit away from circularity, does radiation circularize it back? This question has been the topic of several studies in different scenarios, but never for the system at hand~\cite{German:2023bye,Glampedakis:2002ya,Cardoso:2020iji}. In Appendix~\ref{app:adiabatic_approximation} we consider a charged particle moving in a circular orbit in a generic axisymmetric and stationary spacetime and under the influence of an EM field satisfying the same symmetries. We prove that the inclusion of radiation reaction leads the parameters $E$ and $L$ to evolve in exactly the correct proportion for the particle to remain in a circular orbit. This means that an adiabatic evolution between circular orbits characterized only by their energy is a physically reasonable scenario. It remains to be seen whether generic motion tends to circular motion under radiation reaction; we will not provide any further results in this regard.

%
\vspace{-1 em}
\subsection{Analytical and numerical solution on Teukolsky equation} \label{sec:analytical}
%
As we discussed above, to obtain energy fluxes we have to find solutions to the homogeneous radial equation~\eqref{eq:teukolsky_radial}. This can be done analytically in the low frequency limit, that is, when $\omega M \ll 1$~\cite{Starobinsky:1973aij,Starobinskil:1974nkd,Page:1976df,Cardoso:2020iji,Santos:2024tlt}. The procedure was described in detail for perturbations of the Schwarzschild geometry~\cite{Santos:2024tlt}, and the method for Kerr is identical. 

Since the expressions for the energy flux are very cumbersome, we instead write down the actual solutions to the boundary value problem:
\begin{widetext}
\beq
R^\infty _{\omega \ell m}&=& \frac{i^{\ell+1} 2^{-\ell}\ell \, \Gamma (2 \ell)}{\Gamma (\ell)}\left(r - r_{-} \right) \left(\frac{r - r_+}{r-r_-}\right)^{i\frac{3 a-2 M r_+ \omega 
   }{r_+ - r_-}} \left(\omega 
   \left(r- r_+\right)\right)^{-\ell-1} \,  \label{eq:anres_inf} \\ 
   & \times &\  {_2F_1}\left(\ell+2,\ell+2 i M \omega
   -2 i  \frac{3 a-2 M^2 \omega }{r_+ - r_-}+1;2 (\ell+1);\frac{r_+ - r_-}{r_+-r}\right) \, , \nonumber \\
R^H _{\omega \ell m}&=&  \Delta 
   \left((r_+ - r_-) \frac{r- r_+}{r- r_- }\right)^{i \frac{3 a-2 M r_+
   \omega  }{r_+ - r_-}} \,
   _2F_1\left(1-\ell,\ell+2;2 -2 i M \omega +\frac{2 i \left(3 a-2 M^2 \omega
   \right)}{r_+ - r_-};\frac{r_+-r}{r_+ - r_-}\right) \, , \label{eq:anres_hor}
\eeq
\end{widetext}
where we have introduced $\delta = \sqrt{M^2 - a^2}$. To obtain the energy fluxes, one needs only replace these expressions in Eqs.~\eqref{eq:Zlm}-\eqref{eq:ene_flux_hor}. As stated above, the angular eigenfunctions and eigenvalues are obtained using the Black Hole Perturbation Toolkit~\cite{BHPToolkit}. Since we are in the low frequency limit, these were expanded to first order in $\omega$. 

As mentioned before, we also solved the relevant equations numerically. Our procedure was validated by comparing the results in the absence of magnetic field, $\oc=0$, to results obtained using the Black Hole Perturbation Toolkit's package \texttt{Teukolsky}; we found the two methods were in perfect agreement. 

\section{Results} \label{sec:results}
%
%

\begin{figure}[ht!]
\centering
\includegraphics[width=.5 \textwidth]{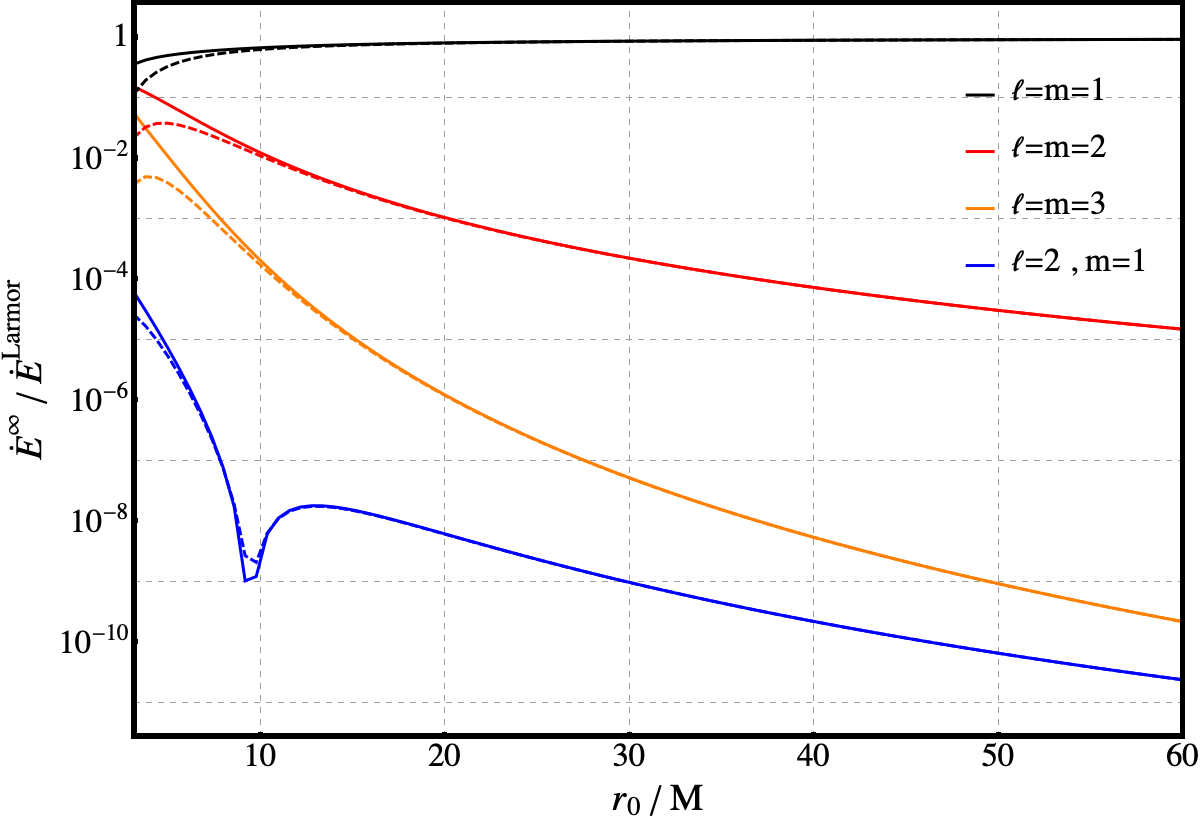}
\caption{\justifying  EM energy flux at infinity, normalized to the Larmor result~\eqref{eq:Larmor}, for a charged particle in a circular prograde plus orbit of radius $r_0$ around a Kerr BH with $a = 0.7 M$ and $M \oc=0.1$. Solid lines are numerical results for different $(\ell,m)$ modes, dashed lines are analytical prediction~\eqref{eq:anres_hor} (we show the sum of flux in the two modes with symmetric azimuthal number $\pm m$, which contribute equally). Lowest orbital radius $r_0$ corresponds to ISCO, $r_\text{ISCO} \approx 3.13 M$. The dipole ($\ell=m=1$) mode is dominant and it recovers the prediction of the Larmor formula for wide orbits with low orbital velocity (see Fig.~\ref{fig:vel_prof}). The other modes are only non negligible near ISCO. The $\ell=2, m=1$ mode exhibits an anti-resonance: for $r_0\approx 10M$ radiation to infinity is suppressed. This phenomenon is also present in higher order modes with $\ell+m$ odd. }
\label{fig:flux_inf}
\end{figure}

\begin{figure}
\centering
\includegraphics[width= .5 \textwidth]{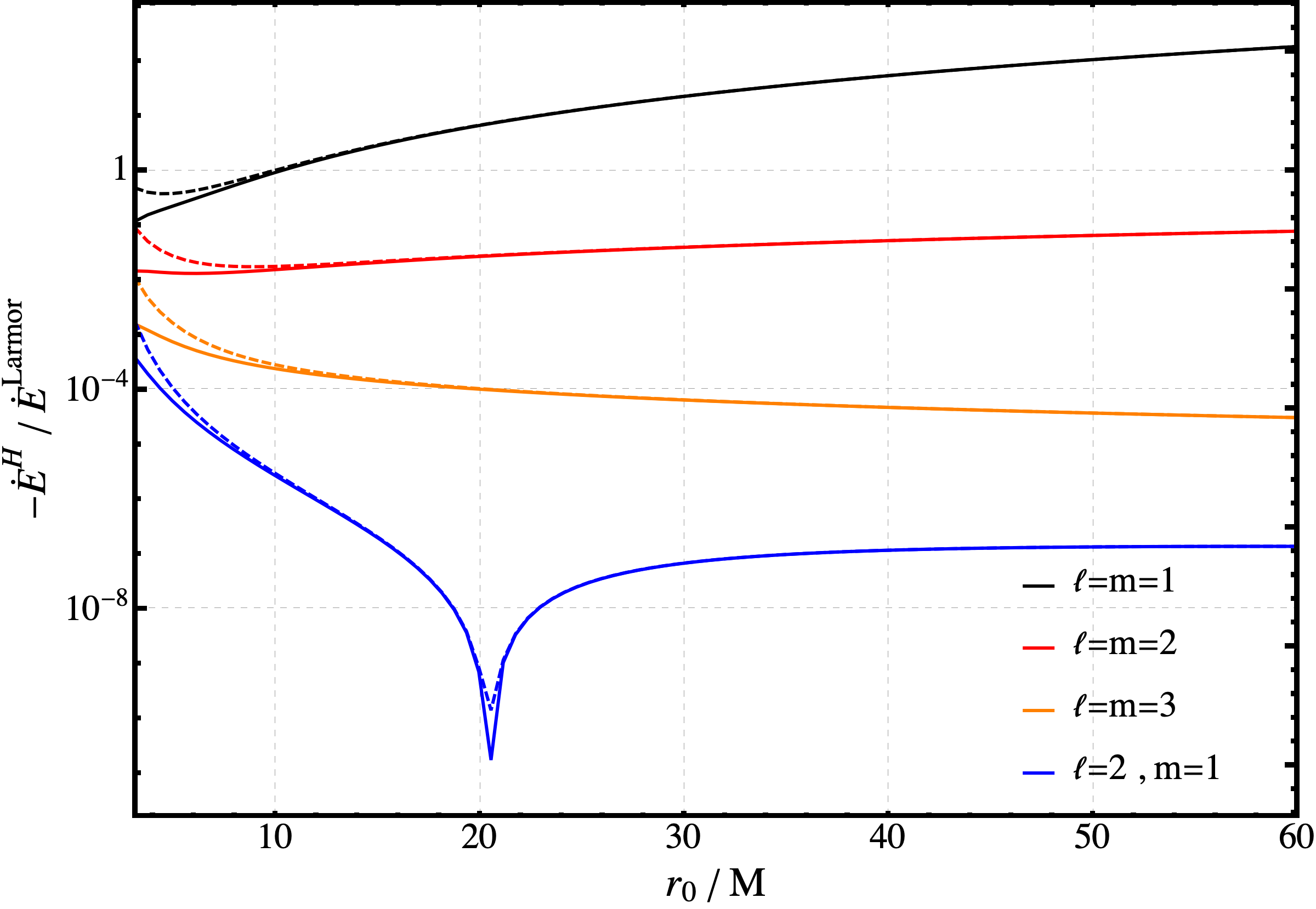}
\caption{\justifying  EM energy flux at the horizon normalized to the Larmor formula~\eqref{eq:Larmor}. Solid lines are numerical results, dashed lines are analytical prediction~\eqref{eq:anres_inf}. 
The flux is negative in all the modes we studied. The dipole dominates and higher order modes can be neglected. The $\ell=2, m=1$ mode exhibits an anti-resonance for $r_0\approx 20M$. }
\label{fig:flux_hor}
\end{figure}

\begin{figure}
\centering
\includegraphics[width= .5 \textwidth]{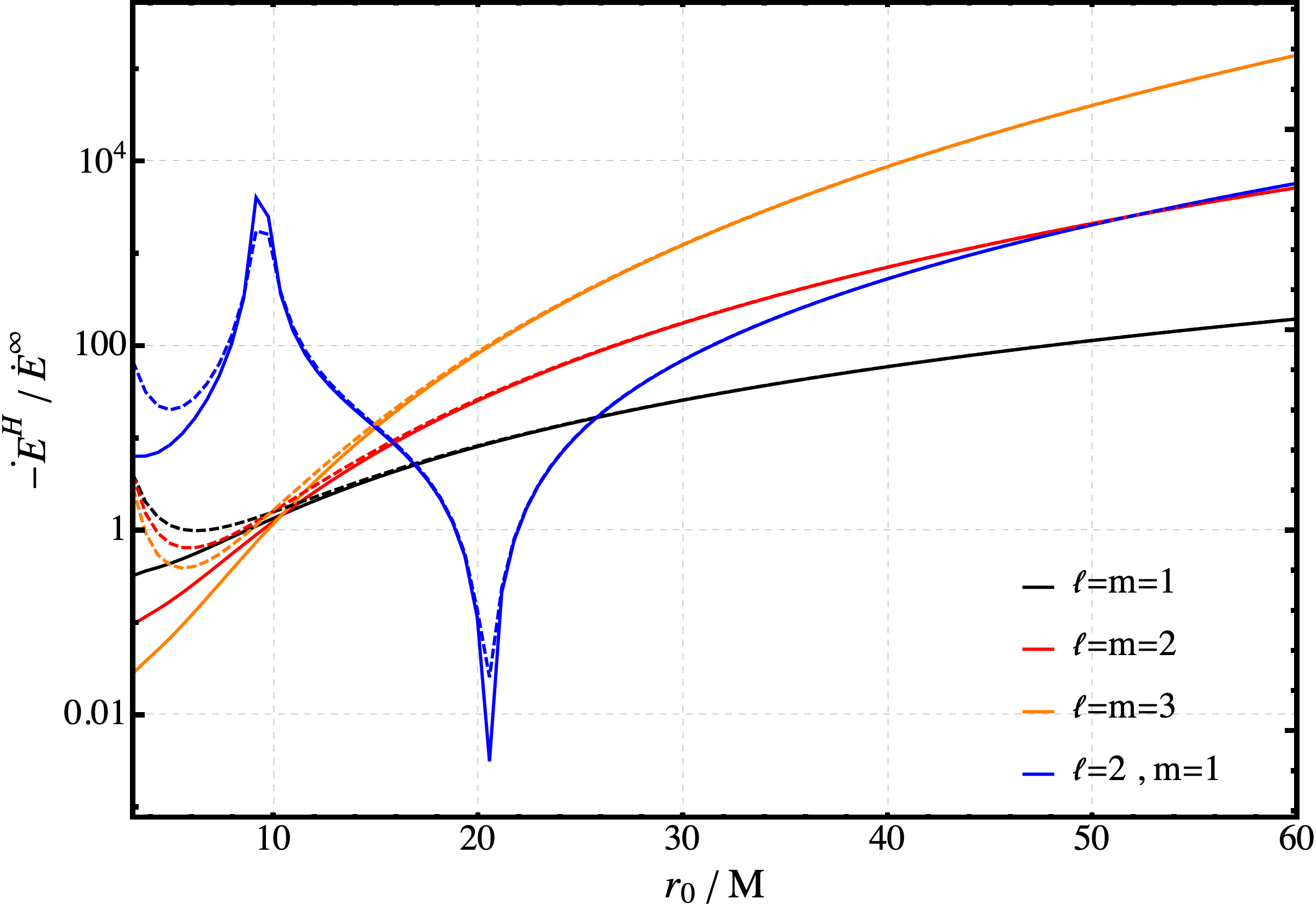}
\caption{\justifying  Ratio of EM energy flux at the horizon and at infinity. Solid (dashed) lines are numerical (low-frequency analytical) results for different $(\ell, m )$ modes. Fluxes at the horizon dominate the energy balance, leading to floating orbits. The anti-resonances of Figs.~\ref{fig:flux_inf}--\ref{fig:flux_hor} are also apparent here.}
\label{fig:flux_rat}
\end{figure}

Our results for EM radiation are summarized in Figs.~\ref{fig:flux_inf}--\ref{fig:flux_rat}. Fluxes at infinity, for the first few multipoles, are shown 
in Fig.~\ref{fig:flux_inf}, normalized to the Larmor result
\begin{equation}
\dot{E}^\text{Larmor} = \frac{2}{3}q^2 r_0^2 \Omega_0^4 \, . \label{eq:Larmor}
\end{equation}
The low-frequency analytical approximation works extremely well in all the range of orbital radii $r_0$. Slight disagreement for $r_0<8 M$ is due to the approximation breaking down, since $M \omega \sim 1$ in this regime.

The radiation is predominantly dipolar; at large $r_0$ we find very good agreement between the Larmor result and the flux in the dipolar mode. Higher order modes can be non-negligible, but only for high frequency orbits inside $r_0<8M$. We can also see that the mode with $\ell=2, \, m=1$ is significantly suppressed when compared with the $\ell=m$ modes. We find this to be a general feature of modes with odd values of $\ell+m$; in addition, the dominant mode satisfies $\ell=
|m|$~\cite{Santos:2024tlt}.

The energy going into the horizon behaves in a similar way, with an exciting twist. Fluxes at the horizon are shown in Fig.~\ref{fig:flux_hor}, normalized to $\dot{E}^\text{Larmor}$, Eq.~\eqref{eq:Larmor}. As we noted already for non-spinning BHs, the energy flux into the horizon can be large, in fact much larger than the energy flux to infinity. This enhancement is apparent in Fig.~\eqref{fig:flux_hor}, which shows that dipolar fluxes {\it into the horizon} can be orders of magnitude larger than those predicted by Larmor's result.

More interesting is the fact that the energy flux is {\it negative} at the horizon, a clear sign that these are superradiant modes~\cite{Brito:2015oca},
and a consequence of the low orbital frequency of the circular orbits we are studying. Condition~\eqref{eq:superradiance} for superradiance to take place is equivalent to requiring that $\Omega_0<\Omega_H \approx 0.2 M^{-1}$. Indeed, this condition is met by all the orbits we are studying (see Fig.~\ref{fig:vel_prof}).

The impact of superradiance on the overall emission process is usually negligible: fluxes at the horizon are suppressed by a high power of the velocity of the orbiting particle~\cite{Poisson:1994yf}. However, this setup is exceptional: radiation fluxes from isolated, charged particles around magnetized BHs {\it can be dominated by the horizon}. This is shown very clearly in Fig.~\ref{fig:flux_rat}, where we plot the ratio between the energy flux at the horizon and at infinity for different modes. 

\begin{figure}
\centering
\includegraphics[width= .48 \textwidth]{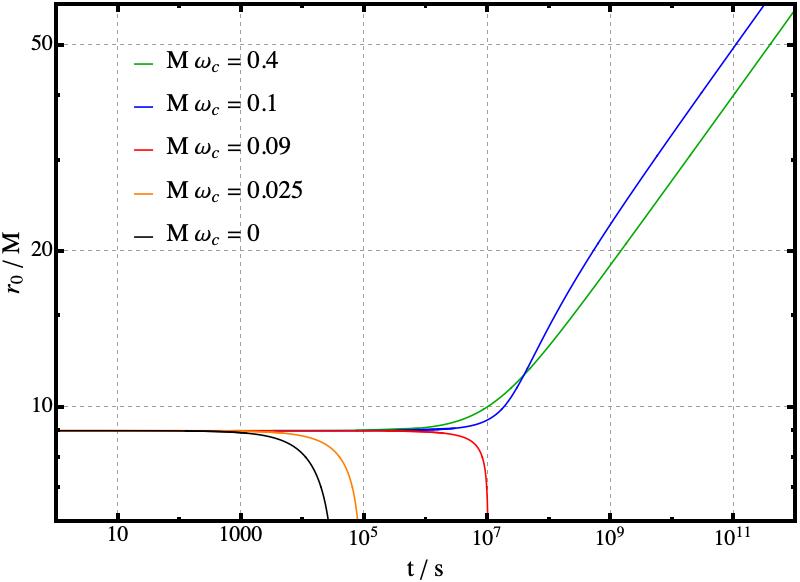}
\caption{\justifying  Adiabatic evolution of orbital radius $r_0$, as a function of the coordinate time $t$, for a radiating charged particle around a magnetized Kerr BH. The particle has charge  $q=1 \, C$ and $m \approx 3.3 \times 10^{-4} \, kg$, and the BH mass is $M= 10 M_{\odot}$. Differently colored curves correspond to different values of the cyclotron frequency $\oc$. The initial orbit is at $r_0 = 9M$, which is slightly larger (smaller) than the radius of the floating orbit for the blue (red) curve. Thus, the particle remains close to the initial orbit for a long time, eventually spiraling out (in). Larger (smaller) values of $M \oc$ than for the blue (red) curve also outspiral (inspiral). Finally, the larger the value of $\oc$ the larger the timescale of the evolution. This point explains why the blue and green curves intersect each other.}
\label{fig:orbit_evo}
\end{figure}

The ratio between the fluxes at the horizon and at infinity in Fig.~\ref{fig:flux_rat} is negative everywhere. In fact, for sufficiently large orbital radius, the energy flux at the horizon becomes larger (in absolute value) than the flux at infinity. For the dominant dipole mode, "sufficiently large" means $r_0 > 9M$. This is closely related to the ``horizon dominance effect" that was found for Schwarzschild BHs in~\cite{Santos:2024tlt}. 

From a dynamical perspective, the evolution of the particle is governed by an energy balance equation,\footnote{Note this equation rests on the assumption that the charges generating the magnetic field do not exchange energy with the system.}
\be
\dot{E}_{b}+\dot{E}^{\rm H}+\dot{E}^{\infty}=0\,,
\ee
where $E_{b}$ is the binding energy of the system. When the flux at the horizon is negligible (or positive), the balance above shows immediately that the particle has to inspiral: since we always have $\dot{E}^{\infty}>0$, to obey the balance equation the system needs to bind more strongly. However, when $\dot{E}^{\rm H}+\dot{E}^{\infty}<0$, as is the case here, the opposite happens: all the energy radiated at infinity is extracted from the BH rather than from binding energy. The particle outspirals rather than inspiralling. In fact, given other {\it dissipation} mechanisms, the particle may float.

Simply put, orbits inside $r_0 \approx 9M$ shrink as a result of radiation reaction, because the particle is losing energy; conversely, orbits outside $r_0 \approx 9M$ widen, since the particle is gaining energy. These later orbits are called \textit{floating orbits}, and they do not exist in the absence of magnetic field~\cite{German:2023bye} (except if there are massive fundamental fields present~\cite{Cardoso:2011xi}). By continuity, there is a particular value of orbital radius $r_0 \approx 9M$ for which the energy coming out of the horizon equals the energy going to infinity. For this value of $r_0$, we have a stationary orbit; the particle remains in the same orbital radius, neither losing nor gaining energy.

To better visualize the floating and outspiraling effects mentioned above, we represent in Fig.~\ref{fig:orbit_evo} some orbital evolutions, using plots of the orbital radius $r_0$ as a function of coordinate time $t$. Concretely, we consider a charged particle with $q=1 \, C$ and $m \approx 3.33\times 10^{-4} \, kg$, moving in the vicinity of a black hole with $M= 10 M_{\odot}$, for various cyclotron frequencies $\oc$. The value of $m$ is chosen so that $M \oc =0.1 \iff B_0 = 10^8 $ Gauss.  The particle starts at $r_0 = 9M$, which is barely outside (inside) the floating orbit for $M \oc=0.1$, corresponding to the blue curve ($M \oc=0.09$, corresponding to the red curve). Thus, for $M \oc \leq 0.09$ we see an inspiral, while the for $M\oc \geq 0.1$ we see an outspiral. Notice how for the curves where $r_0=9M$ is close to being a floating orbit (blue and red) the particle lingers around this value of $r_0$ before spiraling in or out. Finally, the larger $\oc$ the smaller the orbital velocity (see Fig.~\ref{fig:vel_prof}), so the evolution happens on larger timescales. This is why the blue and green curves intersect each other.

Finally, a few words on anti-resonances, i.e., the inverted peaks around $r_0\approx 10M$ and $r_0 \approx 20M$ in Figs.~\ref{fig:flux_inf} and \ref{fig:flux_hor}, respectively. These are single orbits for which the flux at infinity and at the horizon are orders of magnitude lower than for nearby orbits. This effect occurs for orbits with $\ell+m$ odd, at slightly different values of $r_0$ depending on $\ell, m$. Thus, this effect cannot be exclusively governed by the mode frequency $\omega=m\Omega_0$: otherwise, modes sharing the same azimuthal number would have anti-resonances at the same orbit. We have no convincing explanation for this effect, but it is worth noticing that the radiation in these modes is nevertheless suppressed when compared with the dominant the $\ell=m=1$ mode.

\section{Conclusions}\label{sec:conclusions}
Magnetic fields are ubiquitous in nature.
In this work, we studied radiation from charged particles around weakly magnetized BHs.
%
%
We solved the relevant Teukolsy equation describing the EM radiation properties, both numerically and analytically (in the low-frequency regime). Both approaches agree very well in the overlapping regime of validity.
Perhaps unsurprisingly, the dipolar mode dominates radiation emission.
More interestingly, this radiation falls within the superradiant regime, that is, the energy flux at the horizon is negative. For $r_0\gtrapprox 9M$, the energy {\it coming out} of the horizon is larger than that lost at infinity: the particle \emph{extracts energy from the BH} and will spiral outwards, the first example of floating with standard model fundamental fields. Orbits satisfying this condition are called \emph{floating orbits}, and they do not occur in the unmagnetized Kerr BH case.

We expect floating to also occur for gravitational and scalar radiation. This expectation is motivated by the fact that the ``horizon dominance effect" is present in all three channels~\cite{Santos:2024tlt}, and is the driver of the energy extraction mechanism. In astrophysical setups, charged particles will be part of an accretion disk or ionized gas, and friction forces (i.e., nearest neighbhour interactions) are expected to dominate~\cite{Abramowicz:2011xu}, leading to a net energy loss (thus destroying floating). Finally, the extension of our results to non-circular motion is both interesting and important, since it would test the robustness of the energy extraction mechanism. We leave this for future work.

\acknowledgments
%
We thank Rodrigo Vicente, Diogo Ribeiro and Martin Kolo\v{s} for insightful discussions. This work makes use of the Black Hole Perturbation Toolkit.

We acknowledge support by the VILLUM Foundation (grant no.\ VIL37766) and the DNRF Chair program (grant no.\ DNRF162) by the Danish National Research Foundation.
V.C.\ is a Villum Investigator and a DNRF Chair.  
J.S.S.\ and V.C.\ acknowledge financial support provided under the European Union’s H2020 ERC Advanced Grant “Black holes: gravitational engines of discovery” grant agreement no.\ Gravitas–101052587. 
Views and opinions expressed are however those of the author only and do not necessarily reflect those of the European Union or the European Research Council. Neither the European Union nor the granting authority can be held responsible for them.
J.N.\ was partially supported by FCT/Portugal through CAMGSD, IST-ID, projects UIDB/04459/2020 and UIDP/04459/2020.
This work has received funding from the European Union's Horizon 2020 research and innovation programme H2020-MSCA-2022-SE through projects EinsteinWaves, GA no.\ 101131233, and GRU, GA no.\ 101007855.
\vspace{-1 em}
%
%
%
\appendix
\section{Circular orbit evolution}  \label{app:adiabatic_approximation}
%
%
%
In this appendix we show that a charged particle, initially in a circular orbit in the equatorial plane, will radiate energy $E$ and angular momentum $L$ in precisely the correct ratio for it to remain in (quasi-)circular motion. This statement is proved assuming that the adiabatic approximation is valid. As pointed out in Sec.~\ref{sec:comment_adiabatic}, this does not imply that the circular orbit is stable under radiation.

Consider a charged particle of mass $m$ charge $q$ moving in a generic stationary and axisymmetric background spacetime, described by the line element
\begin{equation}
	ds^2 = g_{tt} dt^2 + 2g_{t \phi} dt d\phi + g_{rr}dr^2 + g_{\theta \theta} d \theta ^2 + g_{\phi \phi} d \phi^2 \, ,
\end{equation}
where all the metric components are functions of $r$ and $\theta$ only. We also allow for an EM potential satisfying the same symmetry, $A_\mu=A_\mu (r,\theta)$. This symmetry is encoded in the existence of two Killing vector fields, $X^\mu$ and $Y^\mu$ (see Eq.~\eqref{eq:Killing_vectors}), that commute with the 4-potential. The particle's energy per unit mass $\mE$ and angular momentum per unit mass $\mL$ are still given by Eq.~\eqref{eq:conserved_qtys}. We focus on the case where the motion is confined to the equatorial plane, which means that the values of $\mE$ and $\mL$ completely characterize the trajectory. 

 The Hamiltonian description of this system is identical to that given in the main text, except that now we allow for $g_{\mu\nu}$ and $A_\mu$ to be generic stationary axisymmetric metrics and potentials. This means that Eq.~\eqref{eq:sch_charged_hamiltonian} is not changed, but we rewrite $\mathcal{F}$ as 
\begin{equation}
	\mathcal{F} = 1 + \vec{\Xi}^T \cdot \mathcal{M}_g \cdot \vec{\Xi} \, ,
\end{equation}
where
$$ \vec{\Xi} = \left[\begin{matrix}
   \mE \\
   \mL
  \end{matrix}\right] + \vec{A}\, , \quad \vec{A} = \frac{q}{m} \left[\begin{matrix}
   A_t \\
   -A_\phi 
  \end{matrix}\right]\, , \quad \mathcal{M}_g = \left[\begin{matrix}
   g^{tt} & -g^{t\phi}\\
   -g^{t\phi} & g^{\phi \phi}
  \end{matrix}\right] \, ,$$
and $[g^{\mu \nu}] = [g_{\mu \nu}]^{-1}$. Consider the equations for a circular orbit in the equatorial plane, Eqs.~\eqref{eq:circular_cond}:
\beq
	&\vec{\Xi}^T& \cdot \mathcal{M}_g \cdot \vec{\Xi} = -1 \, , \label{eq:circ_orbit_1}\\
	&\vec{\Xi}^T &\cdot \mathcal{M}^\prime _g \cdot\vec{\Xi} + 2 \, \vec{\Xi}^T \cdot \mathcal{M}_g \cdot\vec{A}^\prime  =0 \, , \label{eq:circ_orbit_2}
\eeq
where primes denote derivatives in $r$.  Suppose that the adiabatic approximation is valid, so that the particle is moving in a sequence of circular orbits described by $\mE$ and $\mL$, which are taken to be slowly varying functions of $r$. This is what we call quasicircular motion. Differentiating Eq.~\eqref{eq:circ_orbit_1} gives
\beq
\vec{\Xi}^T& \cdot \mathcal{M}^\prime _g \cdot \vec{\Xi} + 2 \, \vec{\Xi}^T& \cdot \mathcal{M} _g \cdot \vec{\Xi}^\prime = 0 \, . \label{eq:A5}
\eeq
Taking the difference between Eqs.~\eqref{eq:A5} and~\eqref{eq:circ_orbit_2} yields
\beq
	\vec{\Xi}^T \cdot \mathcal{M}_g \cdot (\vec{\Xi}^\prime - \vec{A}^\prime) = 0 \iff \mE^\prime = \Omega_0 \mL^\prime \, . \label{eq:conserved_qtys_evo}
\eeq
This recovers the result of Eq.~\eqref{eq:flat_momentum_change}, since $\omega = m\Omega_0$ for circular orbits and the adiabatic approximation allows us to swap the time derivatives for radial derivatives. 

In conclusion, \textit{if the particle is in a circular orbit, then the relation~\eqref{eq:conserved_qtys_evo} between the radiated energy and angular momentum is precisely such that the particle remains in quasicircular motion.}
%
%
%
%
%
%
%
%
\end{document}